\documentclass[%
reprint,
runinaddress,
showpacs,preprintnumbers,
 amsmath,amssymb,
 aps,
floatfix,
]{revtex4-1}

\usepackage{graphicx}
\usepackage{dcolumn}
\usepackage{bm}
\usepackage{hyperref}
\usepackage{subfigure}
\usepackage{float}
\usepackage[mathlines]{lineno}

\begin{document}

\preprint{APS/123-QED}

\title{Oscillating Casimir force of trapped Bose gas in electromagnetic field}

\author{Seyit Deniz Han} \email{sdenhan@gmail.com}
\author{Ekrem Aydiner}%
\email{Corresponding Author: ekrem.aydiner@istanbul.edu.tr}
\affiliation{Department of
Physics, Faculty of Science, \.{I}stanbul University,
\.{I}stanbul, 34134, Turkey}%

\date{\today}

\begin{abstract}
In this study, we consider motion of the massive and charged Bose 
gas trapped in electric and magnetic fields between two
parallel plates in the $x-y$ plane that are separated by a distance
$d$ in the $z$ direction. We derive analytical expression of the grand canonical potential of the quantum particles by using of Ketterle and van Druten approximation. By evaluating canonical potential, we obtain Casimir potential and force at the Bose-Einstein condensation temperature $T_{c}$.
We also show that Casimir force oscillates depends on distance $d$ for fixed parameters.
\end{abstract}

\maketitle


\section{Introduction}
\label{intro}

It has been shown by the Casimir \cite{Casimir1948} that quantum mechanical fluctuations of the
electromagnetic field in vacuum between conducting boundaries can leads to an
attractive or repulsive long-range interaction. This effect has been called Casimir effect.
This effect has been experimentally measured in 1997 \cite{Lamoreaux1997} and 1998 \cite{Mohideen1998} which
confirms quantum field theory. It is shown that the force depends on
size, temperature, geometry, surface roughness and electronic
properties of the materials \cite{Milton2001,Mostepanenko1997,Bordag2001}.
The practical applications of Casimir effect is now becoming more widely appreciated
in many fields of physics such as quantum field theory, gravitation and cosmology, Bose-Einstein
condensation (BEC), atomic and molecular systems, mathematical
physics and nano-technology. A good review on new developments in
the Casimir effect can be found in Ref.\,\cite{Bordag2001}.

On the other hand, Fisher and de Genes \cite{Fisher1978} suggested that similar
effect can occur in quantum critical systems. It was so-called as the critical Casimir effect \cite{Krech1994} which is a fluctuation-induced force arise from the confinement of the fluctuations of the order parameter near the critical temperature. Indeed it is shown that the Casimir force in all bosonic systems appears near the critical condensation temperature $T_{c}$ due to thermal fluctuation of these excited states. In recent studies, it is shown theoretically and experimentally that thermal fluctuations of confined quantum
particles in the boundaries can lead to Casimir effect at finite
temperatures with or without traps for the Dirichlet, Neumann and
periodic boundary conditions
\cite{Molina,Pathira,Dai,Dai2,Sisman,Pang,Firat,Nie,Nie2,Nie3,Nie4,Nie5,Dantchev,Martin,Gambassi,Lin,Biswas,Biswas2,Biswas3,Gambassi2,Roberts,Edery,Yu,Napiorkowski,Napiorkowski2,Napiorkowski3,Li,Aydiner,Hasenbusch,Hasenbusch2,Ganshin,Hucht,Vasilyev,Maciolek}.
The understanding behavior of the Casimir force of confined quantum
critical systems and its thermodynamics are becoming important since
potential applications in physics and other areas such as
engineering. Obtained results from theoretical studies may plays
crucial role to invention new electromechanical applications such as
the producing of micro and nano-fabrication materials and its
devices (See Refs.\,\cite{Serry,Chan,Capasso}).

So far Casimir effect has been investigated in many systems
\cite{Molina,Pathira,Dai,Dai2,Sisman,Pang,Firat,Nie,Nie2,Nie3,Nie4,Nie5,Dantchev,Martin,Gambassi,Lin,Biswas,Biswas2,Biswas3,Gambassi2,Roberts,Edery,Yu,Napiorkowski,Napiorkowski2,Napiorkowski3,Li,Aydiner,Hasenbusch,Hasenbusch2,Ganshin,Hucht,Vasilyev,Maciolek},
however, to our knowledge, charged and confined quantum particles
under electric and magnetic field in boundaries has never been
studied yet. However this effect may play important role at low
dimensional systems in many area of physics such as understanding the quantum hall mechanism and thermodynamic of micro and nano mechanical systems.
Therefore, in this study, by inspiring previous studies, we consider
charged and confined boson-like particles under electric and
magnetic fields between two closely spaced conducting plates in the
$x-y$ plane that are separated by a distance $d$ in the $z$
direction. We will obtain analytical expressions for the Casimir potential and
Casimir force based on statistical mechanics.
We have also shown that the oscillation frequency of the Casimir force depends on applied magnetic field. With the best of our knowledge, some of the results are obtained for the first time. These result may lead to new discussions on quantum Hall effect for boson systems \cite{Tao1986,Senthil2013}.

This paper is organized as follows: In section II, we define the
motion of charged particle in crossed electric and magnetic fields
between two closely spaced conducting plates and give the
Hamiltonian equation and its eigenvalues. In section III, starting
from these equations we obtain grand canonical potential of the
system. In section IV, we obtain Casimir potential and force. Finally, in section V, we conclude obtained results.

\section{Charged Particle in an Electromagnetic Field}

We consider helical motion of the massive and charged bosonic 
particles in electric and magnetic fields between two
parallel plates in the $x-y$ plane that are seperated by a distance
$d$ in the $z$ direction. For Dirichlet boundary condition the Hamiltonian of the single particle in an electromagnetic field in the $x-y$ plane is given by
\begin{equation}
\widehat{H}=-\frac{\hbar ^{2}}{2m}\nabla ^{2}-\frac{qB}{2m}\widehat{L}_{z}+%
\frac{q^{2}B^{2}}{8m}(x^{2}+y^{2})+q\varphi
\end{equation}
where $m$ is the mass, $q$ is the charge of particle, $B$ is the external magnetic field, $\varphi$
is the electromagnetic scalar potential. Here, the homogenous magnetic field and electrical field are chosen that as oriented along $z$ and $x$ directions, respectively. It is well known that the charged particles rotate in $x-y$ plane with frequency $w$ depends on parameter of Hamiltonian. The single particle energy levels for Hamiltonian in Eq.\,1  are given \cite{Kubisa1997}
\begin{equation}
\varepsilon _{n_{x,y,z}}=\hbar w_{c}\left[ n_{x}+n_{y}+1\right] +\frac{%
\hbar ^{2}\pi ^{2}}{2md^{2}}n_{z}^{2}+\frac{1}{2}mc^{2}\frac{F^{2}}{B^{2}}-qFW_{0}
\end{equation}
where $w_{c}=-qB/mc$ is the cyclotron frequency, $n_{x,y,z}$ is the Landau
quantum number, $W_{0}=k_{x}L^{2}+\frac{mc^{2}F}{qB^{2}}$ denotes the $y$ coordinate of the
center of oscillations, and $L=(\hbar c/eB)^{1/2}$ is the magnetic radius. The free motion of particle occurs in the $z$ direction
while cyclotron motion of particle occurs in the $x-y$ plane.
In this study, for simplicity we ignore drift motion of particle in the any direction.
Therefore we neglected contribution of $k_{x}$ term in $W_{0}$.

\section{Grand Canonical Potential}

To obtain Casimir force of boson-like particle systems in the statistical mechanics framework must be needed grand canonical potential of the system. Therefore we focus on compute this potential in this section. For boson-like particles, the grand canonical potential can be written as
\begin{equation}
\varphi (T,\mu ,d)=\varphi _{0}+k_{B}T\sum\limits_{n=1}^{\infty }\ln
(1-ze^{-\beta \varepsilon _{n_{x,y,z}}})
\end{equation}
where $k_{B}$ is Boltzmann constant, $T$ is temperature, $\beta=1/k_{B}T$, $z=e^{\beta(\mu+\varepsilon_{0})}$ is the fugacity with ground state energy $\varepsilon_{0}$ and $\varphi _{0}$ is ground state potential of the particle system. 
It is well known that all bosonic systems go to Bose-Einstein condensation at  any critical temperature $T_{c}$. However all particles do not down to the ground state under $T_c$. 

The ground state potential $\varphi _{0}$ does not any contribute to Casimir potential since the Casimir force in these systems mentioned in introduction caused by thermal fluctuation of excited states under the critical temperature $T_c$. Therefore we can neglect ground state
potential $\varphi _{0}$. Hence, under this assumption, the grand canonical potential of the system can be presented as
\begin{equation}
\varphi (T,\mu ,d)=k_{B}T\sum\limits_{n=1}^{\infty
}\sum\limits_{j=1}^{\infty }\frac{z^{j}}{j}e^{-j\beta \varepsilon
_{n_{x,y,z}}}
\end{equation}
By using energy expression in Eq.\,(2), 
the grand canonical potential can be written for charged particles in two dimensional system in the form
\begin{widetext}
\begin{eqnarray}
\varphi (T,\mu ,d)=k_{B}T\sum\limits_{n_{x}=1}^{\infty
}\sum\limits_{n_{y}=1}^{\infty }\sum\limits_{n_{z}=1}^{\infty
}\sum\limits_{j=1}^{\infty }\frac{z^{j}}{j}
\exp \left\{ -j\beta \left(
\hbar w_{c}(n_{x}+n_{y}+1)+ \frac{\hbar ^{2}\pi ^{2}}
{2md^{2}}n_{z}^{2}-\frac{%
1}{2}mc^{2}\frac{F^{2}}{B^{2}}\right) \right\} .
\end{eqnarray}
To eliminate dependence of $n_x$ and $n_y$ Eq.\,(5) can be arranged as
\begin{eqnarray}
\varphi (T,\mu ,d)=k_{B}T\sum\limits_{n_{z}=1}^{\infty
}\sum\limits_{j=1}^{\infty }\frac{\frac{z^{j}}{j}\exp \left\{ -j\beta
\left( \frac{\hbar ^{2}\pi ^{2}}{2md^{2}}n_{z}^{2}\right) \right\} }{\left(
1-\exp \left\{ j\beta \hbar w_{c}\right\} \right) \left( 1-\exp \left\{
j\beta \hbar w_{c}\right\} \right) }
\exp \left\{ j\beta \frac{mc^{2}}{2}%
\frac{F^{2}}{B^{2}}\right\} \exp \left\{ -j\beta \hbar w_{c}\right\}.
\end{eqnarray}
At this point, to simplify expressions in the denominator of Eq.(6) we can use Ketterle and van Druten approximation which is given by  $\left( 1-\exp \left\{ j\beta \hbar w_{c}\right\} \right) =j\beta \hbar w_{c}$. This approximation yields near exact results is used to evaluate the sum over $j$ in the thermodynamic limit $\hbar w\ll k_{B}T$. Hence the grand canonical potential can be rewritten as
\begin{eqnarray}
\varphi (T,\mu ,d)=\frac{(k_{B}T)^{3}}{\hbar ^{2}w_{c}^{2}}\sum\limits_{n=1}^{\infty }\sum\limits_{j=1}^{\infty }\frac{z^{j}%
}{j^{3}}\exp \left\{ -j\beta \left( \frac{\hbar ^{2}\pi ^{2}}{2md^{2}}%
n^{2}\right) \right\}\exp \left\{ -j\beta \left( \frac{mc^{2}}{2}\frac{%
F^{2}}{B^{2}}-\hbar w_{c}\right) \right\}
\end{eqnarray}
where we set $n_{z}=n$ for simplicity. 
This potential function includes surface, bulk and excited states contributions of confined particle system in boundaries in two-dimensional geometry. Therefore, to find contribution of excited states, the potential in Eq.(7) can be decompose to components. 
By using the Jacobi identity \cite{Hunter}
\begin{equation}
\sum\limits_{n=1}^{\infty }e^{-\pi n^{2}b}=\left( \frac{1}{2\sqrt{b}}-%
\frac{1}{2}\right) +\frac{1}{\sqrt{b}}\sum\limits_{n=1}^{\infty }e^{-\pi
n^{2}/b} \ ,
\end{equation}
Eq.(7) can be presented as  
\begin{equation}
\varphi (T,\mu ,d)=\frac{(k_{B}T)^{3}}{\hbar ^{2}w_{c}^{2}}%
\sum\limits_{j=1}^{\infty }\frac{z^{j}}{j^{3}}\left[ \left( \frac{1}{2\sqrt{%
b}}-\frac{1}{2}\right) +\frac{1}{\sqrt{b}}\sum\limits_{n=1}^{\infty
}e^{-\pi n^{2}/b}\right] e^{j\beta A}
\end{equation}
where $b=j\left( \frac{\lambda }{d}\right) ^{2}$ with $b>0$, $A=\frac{mc^{2}}{2}\frac{F^{2}}{B^{2}}-\hbar w_{c}$ and $\lambda =\frac{h}{ \sqrt{2 \pi mk_{B}T}}$ is thermal de Broglie wavelength of the particles. 
\end{widetext}

\section{Casimir Potential and Force}
\label{sec:2}

In Eq.\,(9), the first term corresponds to bulk potential $\varphi_{bulk}\left(T,\mu,d\right)$ and second term is the surface potential $\varphi_{surf}\left(  T,\mu,d\right)$. These terms do not contribute to Casimir potential since they do not include excited states. However, the third term corresponds to thermal fluctuations of excited states of Bose-like gas  which cause to Casimir effect. This term is so-called Casimir potential. The Casimir potential of the trapped boson-like gas in present work is given by
\begin{equation}
\varphi _{C}(T,\mu ,d)=\frac{(k_{B}T)^{3}}{\hbar ^{2}w_{c}^{2}}\frac{d}{%
\lambda }\sum\limits_{j=1}^{\infty }\sum\limits_{n=1}^{\infty }\frac{%
e^{j\beta \mu }}{j^{7/2}}e^{-\frac{\pi }{j}(\frac{nd}{\lambda }%
)^{2}}e^{j\beta A} \ . 
\end{equation}
\small The sum over $j$ which is part of Eq. (10) can be converted to integral form in the limit $d/\xi\ll1$ as
\begin{equation}
\sum\limits_{j=1}^{\infty }\sum\limits_{n=1}^{\infty }\frac{e^{j\beta \mu
}}{j^{7/2}}e^{-\frac{\pi }{j}(\frac{nd}{\lambda })^{2}}e^{j\beta A}=2\left(
\frac{\lambda }{d}\right) ^{5}\sum\limits_{n=1}^{\infty
}\int\limits_{0}^{\infty }x^{-6}e^{-px^{2}-q/x^{2}}dx
\end{equation}
where $p=\frac{u^{2}}{2}=-(\frac{d}{\lambda })^{2}\beta (\mu +A)\sim d/\xi $, $\xi$ is the correlation length. On the other hand, $q=\pi n^{2}$ and $x^{2}=\left( \frac{\lambda }{d}\right) ^{2}j$. By using equality in Eq.\,(11), the Casimir potential is arranged as
\begin{equation}
\varphi _{C}(T,\mu ,d)=\frac{2(k_{B}T)^{3}}{\hbar ^{2}w_{c}^{2}}\left(
\frac{\lambda }{d}\right) ^{4}\sum\limits_{n=1}^{\infty
}\int\limits_{0}^{\infty }x^{-6}e^{-px^{2}-q/x^{2}}dx \ . 
\end{equation}
If the integral in Eq.\,(12) is evaluated \cite{Ryzhik}, the Casimir potential is given by
\begin{widetext} 
\begin{equation}
\varphi _{C}(T,\mu ,d)=-\frac{2(k_{B}T)^{3}}{\hbar ^{2}w_{c}^{2}}\left(
\frac{\lambda }{d}\right) ^{4}\sum\limits_{n=1}^{\infty }\left( \frac{2\pi
u^{2}n+3\sqrt{2\pi}u+ \frac{3}{n}}{8\pi ^{2}n^{4}}\right) e^{-\sqrt{2\pi }un}
\end{equation}
or by using thermal wavelength $\lambda =\frac{h}{ \sqrt{2 \pi mk_{B}T}}$, Eq.\,(13) is rewritten as 
\begin{equation}
\varphi _{C}(T,\mu ,d)=\frac{3k_{B}T\hbar ^{2}}{m^{2}w_{c}^{2}}\frac{1}{%
d^{4}}\sum\limits_{n=1}^{\infty }\left( \frac{\frac{2}{3}\pi u^{2}n^{2}+\sqrt{2\pi}un%
+1}{n^{5}}\right) e^{-\sqrt{2\pi }un} \ .
\end{equation}
where $u=\sqrt{-2(\frac{d}{\lambda })^{2}\beta A}$. This potential caused by two dimensional motion of spinless charged (boson like) particles between plates separated a distance $d$.
\end{widetext} 

Now, we can compute Casimir force corresponds to Casimir potential (14).  At finite temperature $T$, the Casimir force is given by  \cite{Dantchev,Martin}
\begin{equation}
F_{C}(T,\mu ,d)=-\frac{\partial }{\partial d}\left[ \varphi_{C} (T,\mu
,d)-\varphi_{C} (T,\mu ,\infty )\right]
\end{equation}
where $\varphi (T,\mu,d)$ and $\varphi_{C}\left(T,\mu,\infty\right)$ are potential of the system  between the plates and infinite range, respectively. 
It is assumed that in the infinite limit the grand canonical potential goes to zero, i.e., $\varphi_{C}\left(T,\mu,\infty\right)\rightarrow0$. 
Therefore, Casimir force in Eq.\,(15) reduces to 
\begin{equation}
F_{C}(T,\mu ,d)=-\frac{\partial }{\partial d}\varphi _{C}(T,\mu ,d) \ .
\end{equation}
By using Eq.\,(16) Casimir force caused by thermal fluctuation of these excited states under the critical temperature $T_{c}$ can be computed. 
It is well known that the chemical potential $\mu$ is zero at condensation case. When $\mu=0$, it is expected that potential can be simplified easily. However
for $\mu=0$, $u$ does not go to zero, it takes non-zero values  as $u=\sqrt{-2(\frac{d}{\lambda })^{2}\beta A}$.
We must remark that if $u$ had been zero, the contribution to Casimir force would has come from only third term depends on Zeta function $\zeta (5)$ in the parenthesis of Eq.(14). 
But, all terms in Eq.\,(14) give contribution to Casimir force since $u$ takes non zero values for $\mu=0$.  Therefore we investigate contributions of all terms separately follow. Henceforth, all figures have been plotted by setting  Boltzmann's constant $k=1$, speed of light $c=1$, particle mass $m=1$, Planck constant $h=1$, critical temperature $T_{c}=1$ and electric field force $F=1$ for the sake of simplicity.

\subsection{Contribution of the term with order $1/n^5$}

In first case, we focus the behavior of third term of potential (14) neglecting first and second terms in Eq.\,(14). In this case the Casimir potential can be given as
\begin{equation}
\varphi _{C}(T_{c},\mu ,d)=\frac{3k_{B}T_{c}\hbar ^{2}}{m^{2}w_{c}^{2}}\frac{1}{%
d^{4}}\sum\limits_{n=1}^{\infty }\left( \frac{
1}{n^{5}}\right) e^{-\sqrt{2\pi }un} \ .
\end{equation}
To evaluate the summation in Eq.\,(17) we can use polylogarithmic function which is defined by
\begin{eqnarray}
Li_{m}[z]= \sum\limits_{n=1}^{\infty}\frac{z^n}{n^m}=z+\dfrac{z^2}{2^m}+\dfrac{z^3}{3^m}+...
\end{eqnarray}
\begin{figure}[!h]
\centering
\includegraphics[scale=0.315]{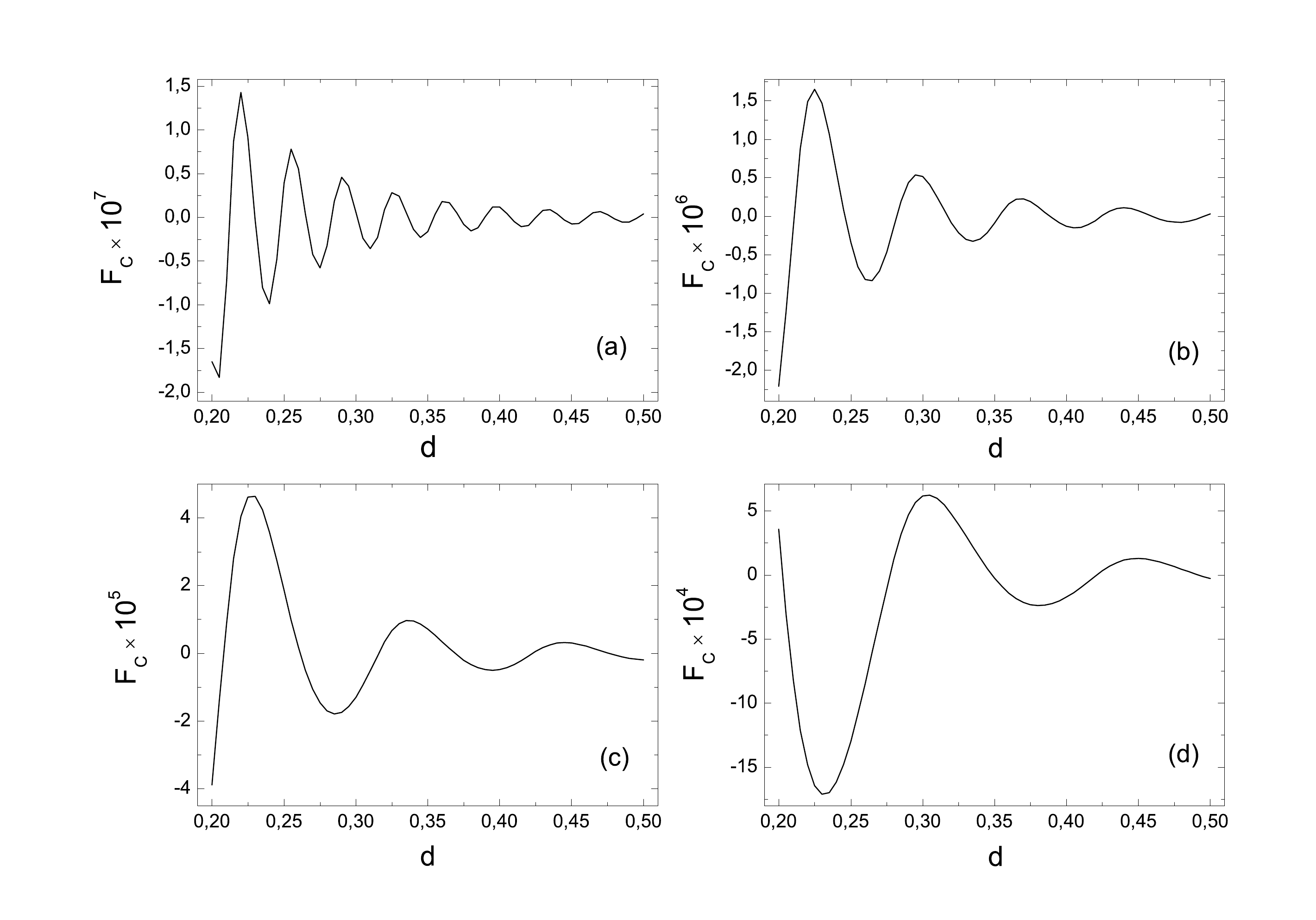}
\caption{The distance $d$ of two parallel plates dependence of the Casimir force $F_{C}$ for set parameters depends on order of $1/n^5$ at (a) $B=0.02$, (b) $B=0.04$, (c) $B=0.06$, (d) $B=0.08$ .}
\label{fig:figure1}
\end{figure}
Comparing Eqs.\,(17) and (18) we can define $m=5$ and $z=\exp(-\sqrt{2\pi }u)$.  By differentiating (17) as to $d$, hence, Casimir force can be found in terms of  
$\operatorname{Li}_{n}[z]$ function as 
\begin{equation}
F_{C}(T_{c},\mu ,d)=\frac{3k_{B}T_{c}h^2}{2m^{2}\pi^{2}w_{c}^{2}}(-\frac{\phi}{2}Li_{4}[e^{\phi}]+2 Li_{5}[e^{\phi}])\frac{1}{d^{5}}
\end{equation} where $\phi=-2\sqrt{-\frac{Ad^{2}m}{h^{2}}}\sqrt{2\pi}$. As it can be seen from Eq.\,(19) Casimir force depends on parameters of the Hamiltonian and $Li_{n}[z]$ functions. However, surprisingly, this force has different dependences of $d$. Additionally this Casimir force is a complex which has real and imaginer parts due to $\phi$ has negative root for positive values of parameters. Therefore, this result is very different from for example Casimir force are caused by free particles or harmonic potential. In these example, Casimir force depends on only distance $d$ inversely with different power. 

In order see how to Casimir force of particles between boundaries at Bose-Einstein critical temperature changes depending distance $d$, we plot real part of $F_{C}(T_{c},0,d)$ versus $d$ for different $B$ values at fixed parameters $k=c=m=h=T_{c}=F=1$. In Fig.\,1 we give plot of Casimir force versus $d$ for (a) $B=0.02$, (b) $B=0.04$, (c) $B=0.06$, (d) $B=0.08$. One can seen from these figures  that Casimir force oscillates and decays depends on distance $d$ for chosen magnetic fields. Figures clearly show that the distance of slabs defines sign and magnitude of Casimir force. This is very interesting and unexpected result. We can explain that the reason of this behavior due to \textit{cosine} dependence in real part of the polylogarithmic function. Particles motion at $x-y$ plane with polar angle $\theta$ provides choosing real part of the polylogarithmic function. Therefore, oscillation of Casimir force caused by particles motion at $x-y$ plane with polar angle $\theta$ since the parameter $A$ in Eq.\,(14) is different from zero as $A=\frac{mc^{2}}{2}\frac{F^{2}}{B^{2}}-\hbar w_{c}$ depends on motion of particles on the polar coordinates when $\mu$ equal to zero. On the other hand, one can see that the amplitude and frequency of oscillation depends on strength of the applied magnetic field. It is seen from figure that amplitude and frequency of Casimir force decreases when the strength of external field is increased. The magnetic field strength and the distance of two parallel plates play significant roles on the formation of Casimir force and its behavior in the system.

\subsection{Contribution of the term with order $1/n^4$}

In second case, we focus the contribution of the second term of potential (14) neglecting first and third terms in Eq.\,(14). In this case the Casimir potential can be given as
\begin{equation}
\varphi _{C}(T_{c},\mu ,d)=\frac{3k_{B}T_{c}\hbar ^{2}}{m^{2}w_{c}^{2}}\frac{1}{%
d^{4}}\sum\limits_{n=1}^{\infty }\left( \frac{\sqrt{2\pi}u
}{n^{4}}\right) e^{-\sqrt{2\pi }un} \ .
\end{equation}
Casimir force corresponds to this potential can be found in terms of $\operatorname{Li}_{n}[z]$ function as 
\begin{equation}
F_{C}(T_{c},\mu ,d)=\frac{3k_{B}T_{c}h^2}{4m^{2}\pi^{2}w_{c}^{2}}\frac{1}{d^{5}}(\phi^2Li_{3}[e^{\phi}]-3Li_{4}[e^{\phi}]) \ .
\end{equation} 
\begin{figure}[!h]
\centering
\includegraphics[scale=0.315]{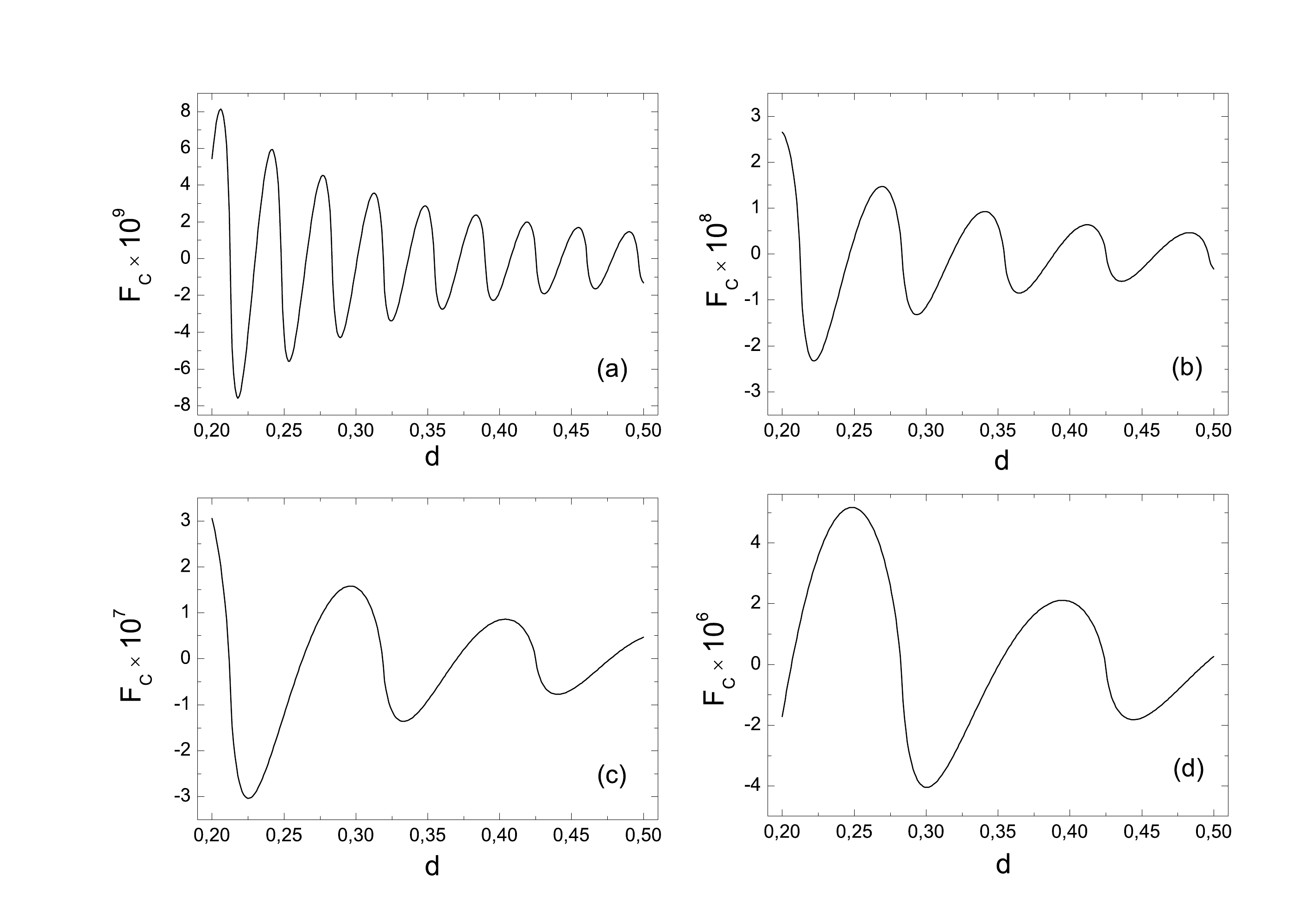}
\caption{The distance $d$ of two parallel plates dependence of the Casimir force $F_{C}$ for set parameters depends on order of $1/n^4$ at (a) $B=0.02$, (b) $B=0.04$, (c) $B=0.06$, (d) $B=0.08$ .}
\label{fig:figure2}
\end{figure}
In order to see contribution to Casimir force of the second term at the condensate phase, we plot real part of $F_{C}(T_{c},0,d)$ in Eq.\,(21) versus $d$ for different $B$ values at fixed parameters $k=c=m=h=T_{c}=F=1$. In Fig.\,2 we give four plots of Casimir force versus $d$ for (a) $B=0.02$, (b) $B=0.04$, (c) $B=0.06$, (d) $B=0.08$. 
Similar oscillating behavior appears in these figures for different magnetic field. The reason of the oscillation is the same mathematical and physical origin. 
Similarly figures clearly show that the distance of slabs defines sign and magnitude of Casimir force. The amplitude and frequency of force decreases for this term when the strength of external field is increased. The main difference from previous case ($1/n^5$) is the $\sqrt{2\pi}u/n$ factor in parenthesis. This difference leads to changing on the amplitude and frequency of Casimir force. Additionally it can be seen from figures that $u=\sqrt{-2(\frac{d}{\lambda })^{2}\beta A}$ in this factor damages oscillation symmetry.

\subsection{Contribution of the term with order $1/n^3$}

In third case, we investigate the contribution of the first term of potential (14) neglecting second and third terms in Eq.\,(14). In this case the Casimir potential can be given as
\begin{equation}
\varphi _{C}(T_{c},\mu ,d)=\frac{3k_{B}T_{c}\hbar ^{2}}{m^{2}w_{c}^{2}}\frac{1}{%
d^{4}}\sum\limits_{n=1}^{\infty }\left( \frac{\frac{2}{3}\pi u^{2}}{n^{3}}\right) e^{-\sqrt{2\pi }un}
\end{equation}
Casimir force corresponds to this potential can be found in terms of $\operatorname{Li}_{n}[z]$ function as 
\begin{equation}
F_{C}(T_{c},\mu ,d)=-\frac{Ak_{B}T_{c}}{m\pi^{3}w_{c}^{2}}\frac{1}{d^{3}}(-\frac{\phi}{2}Li_{2}[e^{\phi}]+Li_{3}[e^{\phi}]) \ .
\end{equation}
In order to see contribution to Casimir force of the second term at the condensate phase, we plot real part of $F_{C}(T_{c},0,d)$ in Eq.\,(23) versus $d$ for different $B$ values at fixed parameters $k=c=m=h=T_{c}=F=1$. In Fig.\,3 we give plot of Casimir force versus $d$ for (a) $B=0.02$, (b) $B=0.04$, (c) $B=0.06$, (d) $B=0.08$. 
Similar oscillating behavior appears in these figures for different magnetic field. The reason of the oscillation is the same mathematical and physical origin. 
Similarly figures clearly show that the distance of slabs defines sign and magnitude of Casimir force. The amplitude and frequency of force decreases for this term when the strength of external field is increased. The main difference from $1/n^5$ is the $2\pi u^{2}/3n^{2}$ factor in parenthesis. This difference leads to changing on the amplitude and frequency of Casimir force. Additionally it can be seen from figures that $u^2=-2(\frac{d}{\lambda })^{2}\beta A$ in this factor strongly violates oscillation symmetry.  
\begin{figure}[!h]
\centering
\includegraphics[scale=0.315]{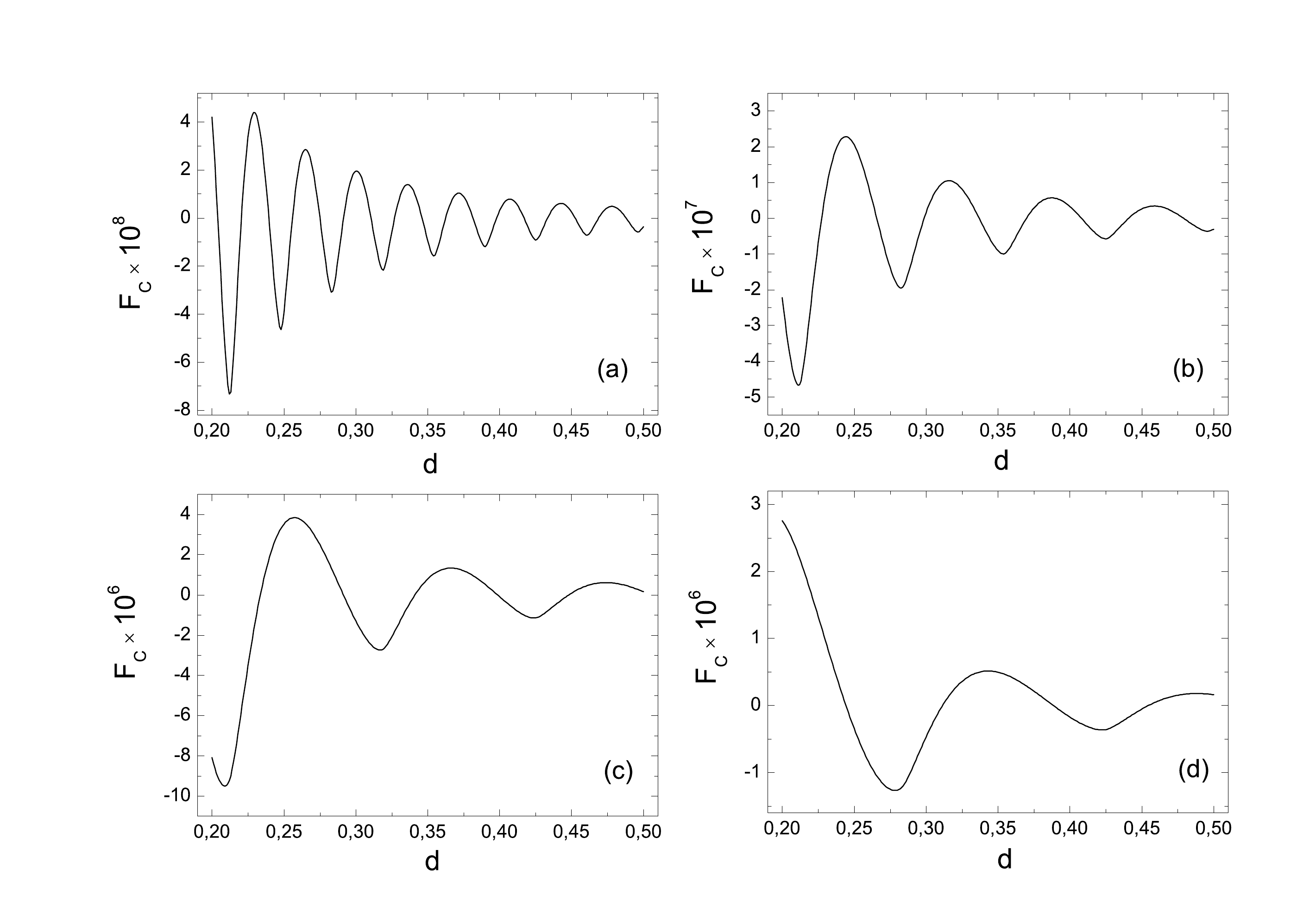}
\caption{The distance $d$ of two parallel plates dependence of the Casimir force $F_{C}$ for set parameters depends on order of $1/n^3$ at (a) $B=0.02$, (b) $B=0.04$, (c) $B=0.06$, (d) $B=0.08$ .}
\label{fig:figure3}
\end{figure}
\begin{figure} [!h]
\centering
\includegraphics[scale=0.315]{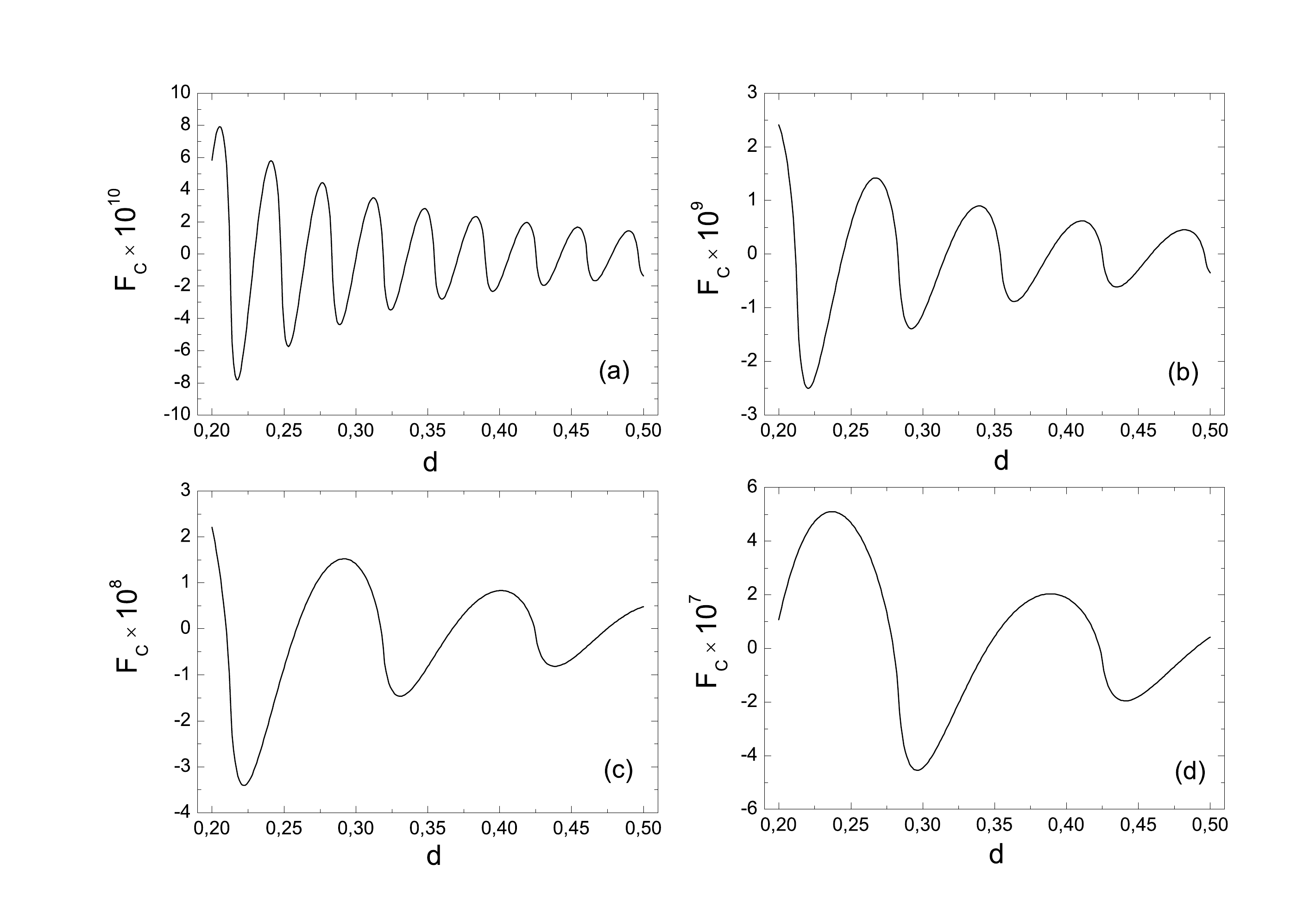}
\caption{The distance $d$ of two parallel plates dependence of the Casimir force $F_{C}$ for set parameters depends on all terms.}
\label{fig:figure4}
\end{figure}

\subsection{Contribution of the all terms}

Finally, if the all terms in Eq.\,(14) are taken into account, the overall Casimir force is obtained as 
\begin{widetext}
\begin{equation}
F_{C}(T_{c},\mu ,d)=-\frac{k_{B}T_{c}}{m^{2}\pi^{2}w_{c}^{2}}\frac{1}{d^{5}}(\phi^{3}\pi^{2}\hbar ^{2}Li_{2}[e^{\phi}]-5\phi^{2}\pi^{2}\hbar ^{2}Li_{3}[e^{\phi}]+12\phi\pi^{2}\hbar ^{2}Li_{4}[e^{\phi}]-12\pi^{2}\hbar ^{2}Li_{5}[e^{\phi}])
\end{equation}
\end{widetext}
In order to obtain contribution of all terms to Casimir force at the condensate phase, we plot real part of $F_{C}(T_{c},0,d)$ in Eq.\,(24) versus $d$ for different $B$ values at fixed parameters $k=c=m=h=T_{c}=F=1$. In Fig.\,4 we give plot of Casimir force versus $d$ for (a) $B=0.02$, (b) $B=0.04$, (c) $B=0.06$, (d) $B=0.08$. Similar oscillating behavior appears in these figures for different magnetic field. The reason of the oscillation is the same mathematical and physical origin. Similarly figures clearly show that the distance of slabs defines sign and magnitude of Casimir force. The amplitude and frequency of force decreases for all term when the strength of external field is increased.This difference leads to changing on the amplitude and frequency of Casimir force. Additionally it can be seen from figures that superposition of all terms strongly violates oscillation symmetry.  

\section{Conclusion}

In this study, we consider motion of the massive and charged bosonic 
particles trapped in electric and magnetic fields between two
parallel plates in the $x-y$ plane that are separated by a distance
$d$ in the $z$ direction. We have derived closed-form expression of the grand canonical potential of the spinless boson-like particles by using of Ketterle and van Druten approximation. By evaluating canonical potential we have obtained Casimir potential and investigated contributions of each terms with different order to Casimir force, separately, at the Bose-Einstein condensation temperature $T_{c}$.

We show that Casimir force of trapped bosonic particles moves in the $x-y$ plane in electric and magnetic fields between two parallel plates oscillates depends on distance $d$ at the Bose-Einstein condensation temperature $T_{c}$. All figures clearly show that the distance of slabs defines sign and magnitude of Casimir force. This interesting and unexpected result may caused from an physical origin. We can explain the reason of this behavior due to \textit{cosine} dependence in real part of the polylogarithmic function. Particles motion at $x-y$ plane with polar angle $\theta$ provides choosing real part of the polylogarithmic function. Therefore, we argue that oscillation of Casimir force caused by particles motion at $x-y$ plane with polar angle $\theta$ because of the parameter $A$ in Eq.\,(14) is different from zero as $A=\frac{mc^{2}}{2}\frac{F^{2}}{B^{2}}-\hbar w_{c}$ depends on motion of particles on the polar coordinates at the critical temperature. 

The detailed investigation of Casimir force for each term with different order and all terms can be useful to setup to implementation and/or verify the oscillating Casimir force experimentally for such kind model. Present investigated model typically likes to two dimensional quantum Hall effect mechanism even though spinless bosonic particles have been considered. However obtained results have potential to lead to new discussions on quantum Hall effect for bosonic systems.

\begin{acknowledgments}
Authors would like to thank Istanbul University for financial support (Grant No. 55383). 
\end{acknowledgments}



\end{document}